\documentclass[twocolumn]{aastex701}

\usepackage{amsmath}
\usepackage{amssymb}
\usepackage{booktabs}     
\usepackage{multirow}

\shorttitle{QSO Composite Spectrum from SPHEREx}
\shortauthors{Kim et al.}

\def\nd{\nodata}
\def\hal{H$\alpha$}
\def\hb{H$\beta$}
\def\paa{Pa$\alpha$}
\def\pab{Pa$\beta$}
\def\pag{Pa$\gamma$}
\def\hei{\ion{He}{1}}

\def\civ{\ion{C}{4}}
\def\mgii{\ion{Mg}{2}}
\def\micron{$\mu{\rm m}$}

\def\farcs{\hbox{$.\mkern-4mu^{\prime\prime}$}}

\begin{document}

\title{A UV-to-Near-infrared QSO Composite Spectrum from the SPHEREx All-Sky Survey}

\author[0000-0002-3560-0781]{Minjin Kim}
\affiliation{Department of Astronomy, Yonsei University, 
50 Yonsei-ro, Seodaemun-gu, Seoul 03722, Republic of Korea}
\email{mkim.astro@yonsei.ac.kr}

\author[0000-0003-1647-3286]{Yongjung Kim}
\affiliation{School of Liberal Studies, Sejong University, 209 Neungdong-ro, Gwangjin-Gu, Seoul 05006, Republic of Korea}
\affiliation{Department of Physics and Astronomy, Sejong University, 209 Neungdong-ro, Gwangjin-Gu, Seoul 05006, Republic of Korea}
\email{}

\author[0000-0002-2770-808X]{Woong-Seob Jeong}
\affiliation{Korea Astronomy and Space Science Institute (KASI) 
776, Daedeok-daero, Yuseong-gu, Daejeon 34055, Republic of Korea}
\email[]{}

\author[0000-0003-3078-2763]{Yujin Yang}
\affiliation{Korea Astronomy and Space Science Institute (KASI) 
776, Daedeok-daero, Yuseong-gu, Daejeon 34055, Republic of Korea}
\email[]{}

\author[0000-0001-9937-8270]{Jeonghyun Pyo}
\affiliation{Korea Astronomy and Space Science Institute (KASI) 
776, Daedeok-daero, Yuseong-gu, Daejeon 34055, Republic of Korea}
\email[]{}

\author[0000-0002-9330-8738]{Richard M. Feder}
\affiliation{University of California at Berkeley, Berkeley, CA 94720, USA}
\email[]{}

\author[0000-0003-1954-5046]{Bomee Lee}
\affiliation{Korea Astronomy and Space Science Institute (KASI) 
776, Daedeok-daero, Yuseong-gu, Daejeon 34055, Republic of Korea}
\email[]{}

\author[0000-0002-6925-4821]{Dohyeong Kim}
\affiliation{Department of Earth Sciences, Pusan National University, Busan 46241, Republic of Korea}
\email[]{}

\author[0000-0002-5037-951X]{Kyuseok Oh}
\affiliation{Korea Astronomy and Space Science Institute (KASI) 
776, Daedeok-daero, Yuseong-gu, Daejeon 34055, Republic of Korea}
\email[]{}

\author[0000-0002-8055-5465]{Jong-Hak Woo}
\affiliation{Department of Physics and Astronomy, Seoul National University, 1 Gwanak-ro, Gwanak-gu, Seoul 08826, Republic of Korea}
\affiliation{SNU Astronomy Research Center, Seoul National University, 1 Gwanak-ro, Gwanak-gu, Seoul 08826, Republic of Korea}
\email[]{}

\author[0000-0003-3301-759X]{Jeong Hwan Lee}
\affil{Research Institute of Basic Sciences, Seoul National University, Seoul 08826, Republic of Korea}
\affil{Department of Physics and Astronomy, Seoul National University, 1 Gwanak-ro, Gwanak-gu, Seoul 08826, Republic of Korea}
\email[]{joungh93@gmail.com}

\author[0000-0002-5437-0504]{Yun-Ting Cheng}
\affil{Department of Physics, California Institute of Technology, 1200 East California Boulevard, Pasadena, CA 91125, USA}
\affil{Jet Propulsion Laboratory, California Institute of Technology, 4800 Oak Grove Drive, Pasadena, CA 91109, USA}
\email[]{}

\author[0000-0001-6320-261X]{Yi-Kuan~Chiang}%
\affiliation{Academia Sinica Institute of Astronomy and Astrophysics (ASIAA), No. 1, Section 4, Roosevelt Road, Taipei 10617, Taiwan}%
\email{ykchiang@asiaa.sinica.edu.tw}%

\author[0000-0002-3892-0190]{Asantha Cooray}%
\affiliation{Department of Physics \& Astronomy, University of California Irvine, Irvine, CA 92697, USA}%
\email{acooray@uci.edu}%

\author[0000-0002-4650-8518]{Brendan~P.~Crill}%
\affiliation{Jet Propulsion Laboratory, California Institute of Technology, 4800 Oak Grove Drive, Pasadena, CA 91109, USA}%
\email{bcrill@jpl.nasa.gov}%

\author[0000-0001-7432-2932]{Olivier~Dor\'{e}}%
\affiliation{Jet Propulsion Laboratory, California Institute of Technology, 4800 Oak Grove Drive, Pasadena, CA 91109, USA}%
\affiliation{Department of Physics, California Institute of Technology, 1200 E. California Boulevard, Pasadena, CA 91125, USA}%
\email{olivier.dore@caltech.edu }%

\author[0000-0002-9382-9832]{Andreas~L.~Faisst}%
\affiliation{IPAC, California Institute of Technology, 770 S. Wilson Ave, Pasadena, CA 91125, USA}%
\email{afaisst@caltech.edu}%

\author[0009-0009-1219-5128]{Zhaoyu Huai}
\affil{Department of Physics, California Institute of Technology, 1200 East California Boulevard, Pasadena, CA 91125, USA}
\email[]{}

\author[0000-0001-5812-1903]{Howard~Hui}%
\affiliation{Department of Physics, California Institute of Technology, 1200 E. California Boulevard, Pasadena, CA 91125, USA}%
\email{hhui@caltech.edu}%

\author[0000-0001-5382-6138]{Daniel C. Masters}
\affil{IPAC, California Institute of Technology, MC 100-22, 1200 East California Boulevard, Pasadena, CA 91125, USA}
\email[]{}

\author[0000-0001-9368-3186]{Chi~Nguyen}%
\affiliation{Department of Physics, California Institute of Technology, 1200 E. California Boulevard, Pasadena, CA 91125, USA}%
\email{chnguyen@caltech.edu}%

\author[0000-0001-8253-1451]{Michael~Zemcov}%
\affiliation{School of Physics and Astronomy, Rochester Institute of Technology, 1 Lomb Memorial Dr., Rochester, NY 14623, USA}%
\affiliation{Jet Propulsion Laboratory, California Institute of Technology, 4800 Oak Grove Drive, Pasadena, CA 91109, USA}%
\email{mbzsps@rit.edu}

\author[0000-0001-5929-4187]{Tzu-Ching~Chang}%
\affiliation{Jet Propulsion Laboratory, California Institute of Technology, 4800 Oak Grove Drive, Pasadena, CA 91109, USA}%
\affiliation{Department of Physics, California Institute of Technology, 1200 E. California Boulevard, Pasadena, CA 91125, USA}%
\email{tzu@caltech.edu}%

\correspondingauthor{Minjin Kim; Yongjung Kim}
\email{mkim.astro@yonsei.ac.kr; yongjungkim@sejong.ac.kr}

\begin{abstract}
We present a composite spectrum of quasi-stellar objects (QSOs) constructed from spectrophotometric data obtained by SPHEREx. The SPHEREx mission surveys the entire sky in 102 near-infrared spectral channels spanning $0.75-5.0$~\micron\ with a spectral resolution of $R \approx 35-130$, providing a unique dataset for building a statistically robust QSO composite from the rest-frame ultraviolet to the near-infrared. Our sample comprises $\sim 61,000$ type 1 QSOs drawn from the Sloan Digital Sky Survey (SDSS) with spectroscopic redshifts (median $z \approx 1.26$), covering a rest-frame wavelength range of $0.14-4.5$~\micron. We find that the UV and optical continuum of the resulting composite can be described by a power law, $f_\nu \propto \nu^{\alpha_\nu}$, with a best-fit spectral index of $\alpha_\nu = -0.10$, while the near-infrared continuum is well-fit with a spectral index of $-1.46$. The power-law indices in both the optical and near-infrared regimes strongly depend on properties of active galactic nuclei (AGNs), such that more luminous AGNs tend to exhibit flatter UV/optical and steeper near-infrared continua compared to those of less luminous ones. The IR-to-optical flux ratio decreases with increasing AGN luminosity, consistent with the predictions of the receding torus model. Broad emission lines, including \hal, \pab, and \paa, are clearly detected in the composite. Their line ratios are in good agreement with predictions from Case~B recombination, suggesting that internal extinction is almost negligible. The equivalent widths of these emission lines are proportional to AGN luminosity, contrary to the trend expected from the Baldwin effect. Finally, the shape of the composite is sensitive to host-galaxy contamination, which must be considered when utilizing this QSO composite for subsequent scientific applications.
\end{abstract}


\section{Introduction}
\label{sec:intro}
The demographics of Quasi-stellar objects (QSOs) are of fundamental importance for understanding the history of the universe. QSOs are powered by accretion onto supermassive black holes (SMBHs), which drives strong UV and optical emission from the accretion disk. This radiation photoionizes the dense clouds in the broad-line regions (BLRs) and heats the surrounding dusty torus, producing the characteristic broad emission lines and infrared (IR) continuum excess that define the QSO spectral energy distribution (SED). These electromagnetic signatures, a blue featureless continuum, prominent broad emission lines, and a rising IR continuum, have been widely exploited to identify QSOs across diverse survey datasets \cite[e.g.,][]{richards_2002, stern_2012}.

QSOs play a key role in regulating the growth of massive galaxies through energetic feedback, both in kinetic and radiative modes \citep[e.g.,][]{silk_1998, fabian_2012, kormendy_2013}. They may be an important source of ionizing photons during the epoch of reionization \citep[e.g.,][]{fan_2006, kimy_2015, robertson_2015, davies_2018, madau_2024, dayal_2025}, although their relative contribution remains debated. Furthermore, due to their high luminosities and strong emission lines, QSOs serve as powerful tracers of large-scale structure across cosmic time \citep[e.g.,][]{hou_2021,neveux_2020,desi_2025}. Motivated by these applications, numerous QSO search programs have been conducted using large photometric and spectroscopic survey datasets \citep[e.g.,][]{richards_2002, ross_2012, lyke_2020, chaussidon_2023, yang_2023, fu_2024, storey-fisher_2024}. Among the most commonly used selection tools is the color-color diagram, which exploits the distinctive broadband colors of QSOs relative to stars and galaxies. Achieving high completeness and purity in such color-based selections requires accurate knowledge of the QSO SED across the full UV-to-IR wavelength range.

In this context, composite spectra, constructed by co-adding large samples of individual QSO spectra, are invaluable tools. They provide high signal-to-noise ratio (S/N) templates that reveal average spectral properties, enable measurement of faint emission features undetectable in individual objects, and serve as empirical templates for SED modeling and photometric redshift estimation \citep{francis_1991, vandenberk_2001, richards_2006, lusso_2015, selsing_2016}. The composite of \citet{vandenberk_2001}, constructed from over 2,000 Sloan Digital Sky Survey (SDSS) QSO spectra, defined the canonical UV--optical QSO template and has been widely used for over two decades for various purposes. More recently, near-infrared (NIR) composites have been produced using smaller samples \citep{glikman_2006, kim_2015, hernan-Caballe_2016}, extending coverage into the NIR bands where emission from the broad Paschen lines and hot dust becomes prominent. Therefore, combining UV/optical and NIR composites is essential for obtaining a comprehensive view of QSO physics.

The SPHEREx mission, which has recently started surveying, will provide a unique opportunity to explore the QSO population. The all-sky survey spectroscopic dataset covers the optical and NIR wavelengths from 0.75 to 5.0 $\mu$m but with modest spectral resolving power ($R \approx 35-130$, \citealp{bock_2026}). This dataset will also open a new window for investigating a statistically meaningful UV-to-NIR QSO composite, enabled by its broad wavelength coverage and all-sky survey. In this study, we construct, for the first time, a QSO composite that simultaneously covers the UV and NIR spectral regimes, using more than 60,000 sources from the early SPHEREx dataset. Throughout this paper, we adopt a cosmology with $H_0=70$ km s$^{-1}$ Mpc$^{-1}$, $\Omega_m=0.3$, and $\Omega_\Lambda=0.7$, when the luminosity is estimated.

\section{Sample and Data}
\label{sec:data}

\subsection{QSO Sample}
\label{subsec:sample}
To construct a representative QSO composite using the SPHEREx dataset, we require a large sample of QSOs with spectroscopic redshifts and physical property estimates. To this end, we utilize the QSO catalog from SDSS data release 16 \citep[DR16;][]{lyke_2020}. In this catalog, the unobscured QSO candidates were selected based on their optical colors, which exhibit a blue continuum and broad emission lines originating from accretion disks and dense clouds in the BLR, respectively \citep{richards_2002}. The final QSOs were selected from multi-fiber spectroscopic follow-up conducted by BOSS and SDSS spectrographs, during which redshifts were determined. This process resulted in 750,414 objects. 

For those objects, the spectral properties (e.g., the host contamination, fluxes of emission lines, and flux densities in the continuum from the accretion disk) measured from the SDSS optical spectra are provided by \citet{wu_2022}. Based on these measurements, the black hole (BH) mass ($M_{\rm BH}$) has been estimated using a single-epoch virial mass estimator \cite[e.g.,][]{vestergaard_2002, mclure_2002, greene_2005}. Motivated by the correlation between the size of BLR and continuum luminosity of active galactic nucleus (AGN) \cite[e.g.,][]{kaspi_2000}, this method utilizes the full width at half maximum (FWHM) of H$\beta$, \mgii, and \civ\ in combination with corresponding continuum luminosities at 5100, 3000, 1350 \AA\ with the recipe adopted from \cite{shen_2011, shen_2019}. The bolometric luminosity ($L_{\rm bol}$) is calculated using the rest-frame continuum luminosities at 5100, 3000, and 1350 \AA\ using the bolometric corrections of 9.26, 5.15, and 3.81, respectively (\citealp{richards_2006}). The fiducial values of BH mass and bolometric luminosity for each object are determined according to their redshifts (see \citealp{wu_2022} for details). 

The SPHEREx all-sky survey data provides the spectral data in $\sim 102$ channels. We cross-match the QSO catalog with the SPHEREx data obtained from April 24, 2025, to August 21, 2025, and find that 648,845 objects are observed. As the SDSS QSO catalog includes objects down to a limiting magnitude of $g < 22$ mag, not all observed targets can have sufficiently high S/N in the SPHEREx data, which could degrade the quality of the composite spectrum. Therefore, we discard objects with low-quality SPHEREx data, defined as a median S/N per spectral channel less than 2. This selection criterion significantly reduces the sample to $\sim22\%$ of its original size. 
To provide sufficient spectral coverage, we require the object to have spectral measurements in at least 50 distinct spectral channels, as the full-sky survey was incomplete when the data were taken. In addition, a robust spectroscopic redshift is crucial for reliably generating the composite in the rest frame, thereby imposing additional selection criteria on redshift measurements (i.e., \texttt{ZWARNING}=0). 

Finally, because broad absorption features and contamination from the host galaxy can introduce non-linear effects during composite generation, we only include objects without broad absorption line signatures (${\texttt{BAL\_PROB}} = 0$) and with negligible host-galaxy flux contributions ( $f_{\text{host}}$) at 5100 $\text{\AA}$, as estimated from the SDSS spectra \citep{wu_2022}. In particular, we require $f_{\text{host}} < 0.1$. \citet{ren_2024} estimated $f_{\text{host}}$ by fitting SDSS QSO spectra with quasar and host galaxy templates derived via principal component analysis. By adopting a penalized pixel-fitting method \citep[e.g.,][]{merritt_1997} rather than a linear decomposition, the reliability of the host galaxy contribution improved significantly over previous efforts, especially in low S/N regimes \citep{ren_2024}. We note that while we attempt to minimize host galaxy contamination using the SDSS optical spectra, the host contribution remains significant in the SPHEREx composite. This contribution becomes increasingly prominent at longer wavelengths. These selection criteria yield a final sample of 60,954 QSOs.

\begin{figure}[t!]
\centering
\includegraphics[width=0.45\textwidth]{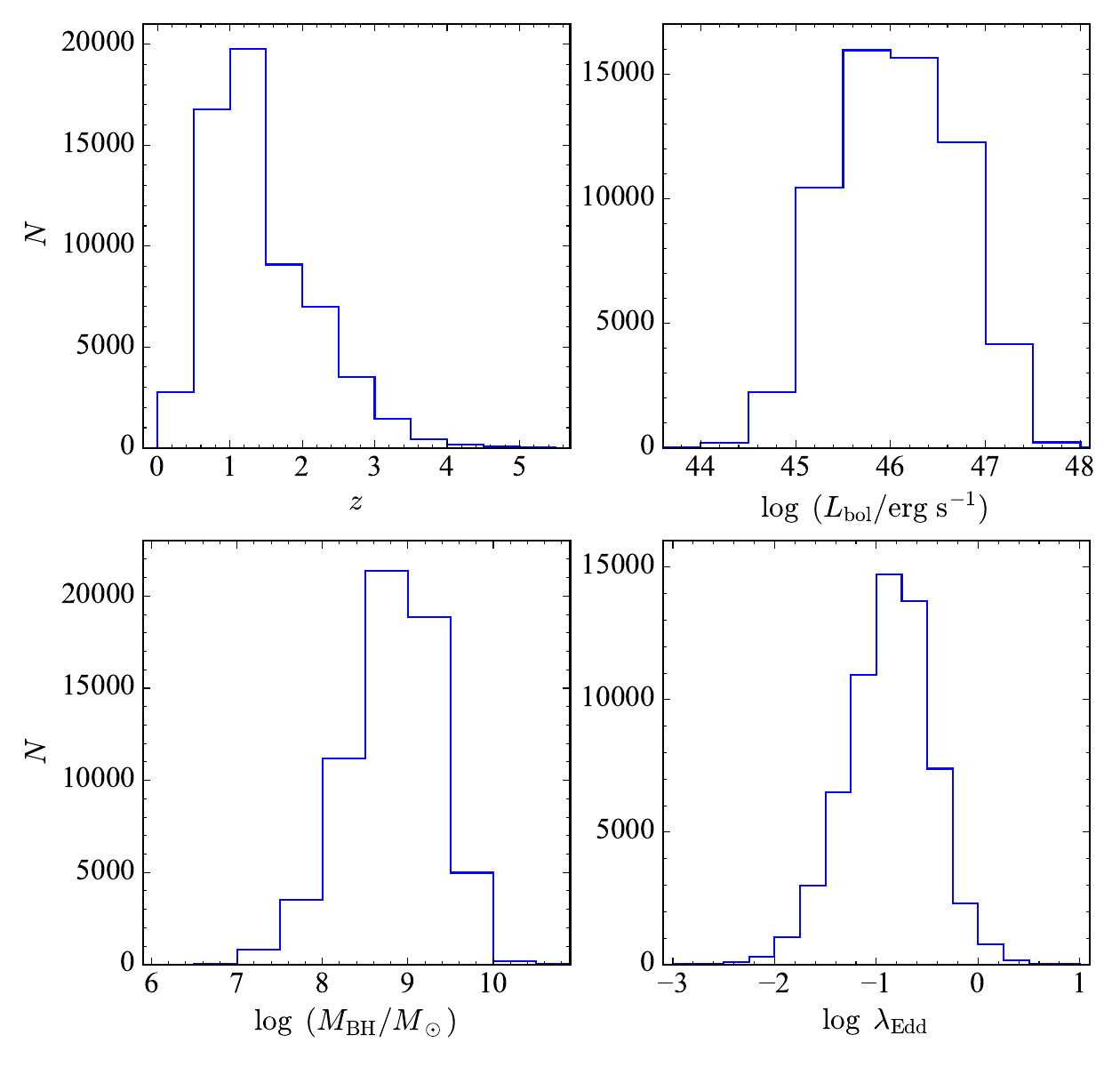}
\caption{
Physical properties of the final QSO sample, used to generate the composite spectra. Redshifts and bolometric luminosities are shown in the top panels, while BH mass and Eddington ratios are displayed in the bottom panels.  
\label{fig:fig1}}
\end{figure}

The physical properties of the final sample are summarized in Table~\ref{tab:sample} and illustrated in Figure~\ref{fig:fig1}. The full sample spans median redshifts $z = 1.26^{+1.02}_{-0.55}$, bolometric luminosities $\log~(L_{\rm bol}/{\rm erg~s^{-1}}) = 46.05^{+0.76}_{-0.71}$, BH masses $\log~(M_{\rm BH}/M_\odot) = 8.86^{+0.52}_{-0.62}$, and Eddington ratios $\log \lambda_{\rm Edd} = -0.85^{+0.40}_{-0.48}$ (median with 16th--84th percentile
ranges). To construct composite spectra for subsets based on the AGN properties,  we divide the full sample into five equal-size ($N \approx 12{,}200$) bins in $L_{\rm bol}$ (bol0--bol4), $M_{\rm BH}$ (bh0--bh4), and Eddington ratio ($\lambda_{\rm Edd}$; edd0--edd4). The properties of each sub-sample are listed in Table~\ref{tab:sample}.

\begin{deluxetable}{cccccc}
\tablecaption{QSO sample properties.\label{tab:sample}}
\tablewidth{\columnwidth}
\tablehead{
    \colhead{Sample} & \colhead{$z$} & \colhead{$\log\ L_{\rm bol}$} & \colhead{$\log\ M_{\rm BH}$} & \colhead{$\log \lambda_{\rm Edd}$} & \colhead{$N$} \\
    \colhead{(1)} & \colhead{(2)} & \colhead{(3)} & \colhead{(4)} & \colhead{(5)} & \colhead{(6)}
}
\startdata
All & $1.26^{+0.95}_{-0.52}$ & $46.05^{+0.71}_{-0.66}$ & $8.86^{+0.48}_{-0.56}$ & $-0.85^{+0.37}_{-0.44}$ & 60954 \\
bol0 & $0.69^{+0.21}_{-0.23}$ & $45.23^{+0.17}_{-0.28}$ & $8.28^{+0.41}_{-0.51}$ & $-1.19^{+0.46}_{-0.42}$ & 12129 \\
bol1 & $0.98^{+0.26}_{-0.26}$ & $45.68^{+0.13}_{-0.14}$ & $8.61^{+0.34}_{-0.39}$ & $-1.04^{+0.38}_{-0.34}$ & 12095 \\
bol2 & $1.27^{+0.25}_{-0.24}$ & $46.04^{+0.12}_{-0.12}$ & $8.84^{+0.31}_{-0.33}$ & $-0.91^{+0.33}_{-0.31}$ & 12173 \\
bol3 & $1.62^{+0.55}_{-0.37}$ & $46.43^{+0.16}_{-0.15}$ & $9.09^{+0.27}_{-0.31}$ & $-0.76^{+0.30}_{-0.27}$ & 12287 \\
bol4 & $2.35^{+0.63}_{-0.53}$ & $46.91^{+0.25}_{-0.17}$ & $9.38^{+0.26}_{-0.32}$ & $-0.55^{+0.30}_{-0.23}$ & 12269 \\
bh0 & $0.74^{+0.36}_{-0.25}$ & $45.37^{+0.43}_{-0.36}$ & $8.12^{+0.20}_{-0.38}$ & $-0.75^{+0.36}_{-0.38}$ & 12090 \\
bh1 & $1.03^{+0.40}_{-0.31}$ & $45.75^{+0.42}_{-0.41}$ & $8.57^{+0.10}_{-0.12}$ & $-0.94^{+0.42}_{-0.40}$ & 12145 \\
bh2 & $1.26^{+0.58}_{-0.38}$ & $46.05^{+0.44}_{-0.43}$ & $8.86^{+0.09}_{-0.09}$ & $-0.92^{+0.42}_{-0.42}$ & 12218 \\
bh3 & $1.53^{+0.77}_{-0.44}$ & $46.38^{+0.40}_{-0.46}$ & $9.13^{+0.10}_{-0.09}$ & $-0.86^{+0.38}_{-0.45}$ & 12254 \\
bh4 & $2.05^{+0.70}_{-0.67}$ & $46.81^{+0.33}_{-0.49}$ & $9.46^{+0.21}_{-0.13}$ & $-0.79^{+0.26}_{-0.49}$ & 12248 \\
edd0 & $0.91^{+0.42}_{-0.29}$ & $45.51^{+0.48}_{-0.42}$ & $8.88^{+0.44}_{-0.41}$ & $-1.42^{+0.15}_{-0.27}$ & 12155 \\
edd1 & $1.12^{+0.44}_{-0.39}$ & $45.85^{+0.45}_{-0.48}$ & $8.82^{+0.44}_{-0.47}$ & $-1.07^{+0.08}_{-0.09}$ & 12147 \\
edd2 & $1.31^{+0.66}_{-0.52}$ & $46.13^{+0.51}_{-0.58}$ & $8.87^{+0.50}_{-0.57}$ & $-0.85^{+0.07}_{-0.07}$ & 12170 \\
edd3 & $1.53^{+0.90}_{-0.66}$ & $46.42^{+0.49}_{-0.72}$ & $8.96^{+0.49}_{-0.72}$ & $-0.65^{+0.08}_{-0.07}$ & 12235 \\
edd4 & $1.93^{+0.85}_{-0.97}$ & $46.63^{+0.45}_{-0.80}$ & $8.80^{+0.50}_{-0.74}$ & $-0.38^{+0.22}_{-0.12}$ & 12247 \\
\enddata
\tablecomments{
Col. (1): Definition of AGN subsamples. Subsamples `bol0'--`bol4', `bh0'--`bh4', and `edd0'--`edd4' are binned by bolometric luminosity, black hole mass, and Eddington ratio, respectively, in increasing order.  
Col. (2): Redshift.
Col. (3): Logarithmic bolometric luminosity,  $\log~(L_{\text{bol}} / {\rm erg~s}^{-1})$.
Col. (4): Logarithmic BH mass, $\log\ (M_{\rm BH}/M_{\odot})$.
Col. (5): Logarithmic Eddington ratio.
Col. (6): Number of objects in each subsample. 
}
\end{deluxetable}

\subsection{SPHEREx Survey Data}
\label{subsec:spherex_data}
The SPHEREx all-sky survey is conducted with six linear variable filters (LVFs), whose bandpass central wavelengths vary along the pixel axis. Spectral images, in which wavelength information is encoded in pixel positions, are obtained at each observation with a spatial resolution of 6\farcs 15; repeated observations with offsets in the wavelength direction enable the construction of full spectral data for specific targets. While SPHEREx covers $3.5^\circ \times 11.3^\circ$ in a single exposure, a dichroic beam splitter divides the light into two channels: $0.75-2.42~\mu\text{m}$ (Bands 1--3) and $2.42-5.0~\mu\text{m}$ (Bands 4--6; \citealp{korngut_2026}).

We utilize photometric data from the SPHEREx all-sky survey. Forced photometry of the LVF images, calibrated for basic reduction (e.g., dark current, flat-field, and photometric corrections; \citealp{akeson_2025}), is performed via point spread function fitting at the positions of known objects in the SPHEREx Reference Catalog (Y. Yang et al., in prep.), which includes all SDSS DR16 QSOs. Ideally, a specific target is observed $\sim102$ times through the scanning, yielding flux density measurements across $\sim 102$ contiguous LVF channels from $0.75$ to $5.0~\mu{\rm m}$. However, because spectroscopic imaging data are collected by scanning the entire sky every six months, the spectral coverage for individual targets may not be uniform, and some targets may be only partially observed. Consequently, we only include targets with sufficient spectral sampling, defined as having more than 50 spectral elements. While the post-processing pipeline provides photometric data interpolated to the 102 fiducial spectral channels, we utilize the native photometric data estimated from the raw continuous sampling in this study. The spectral resolution varies from $R \approx 35 - 41$ in the shortest-wavelength channels ($\le 3.82~\mu\text{m}$) to $R \approx 110-130$ at the longest wavelengths ($> 3.82~\mu\text{m}$; \citealp{bock_2026}). 

To construct a composite spectrum from the SPHEREx dataset, we first remove spurious measurements around $2.42~\mu\mathrm{m}$ caused by the dichroic beamsplitter transition\footnote{Detailed modeling of this wavelength region will be implemented in the new SPHEREx pipeline.} \citep{hui_2026}. We apply a Galactic extinction correction using the dust maps from \citet{schlegel_1998}, alongside the recalibration from \citet{schlafly_2011}, and the extinction law with $R_V = 3.1$ adopted from \citet{gordon_2023}. Finally, the spectral data are converted to the rest frame using redshift measurements from the SDSS optical spectra \citep{lyke_2020}.  Although we utilize cumulative data through 2025 August 21, the current sample provides a statistically significant dataset for producing a representative composite and for comparing subsets of the QSO population.

\begin{figure*}[ht!]
\centering
\includegraphics[width=0.95\textwidth]{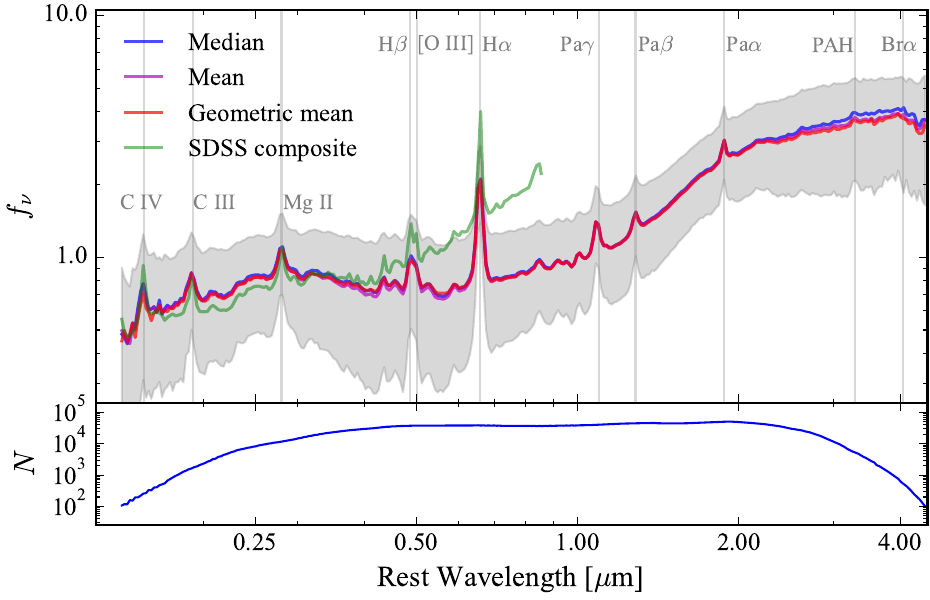}
\caption{
Composite spectra generated from the SPHEREx dataset. For the top panel, blue, magenta, and red lines denote the composites with median, mean, and geometric mean, respectively. Shaded regions indicate the 16th–84th percentile range of the flux distribution in each spectral bin. Vertical gray lines indicate the rest-frame wavelengths of major emission lines, as labeled at the top of the plot. The green line represents the composite QSO spectrum from the SDSS (\citealt{vandenberk_2001}). The SDSS composite is rebinned in wavelength to match the spectral bins of the SPHEREx composite and normalized in the $0.3–0.4$ \micron\ range relative to the median SPHEREx composite. In the bottom panel, the number of samples per spectral bin is shown.  
\label{fig:fig2}}
\end{figure*}

\begin{deluxetable}{ccccc}
\label{tab:composite}
\tablecaption{SPHEREx QSO Composite Spectra.\label{tab:composite}}
\tablewidth{\columnwidth}
\tablehead{
    \colhead{$\lambda_{\rm rest}$} & \colhead{$f_\nu$ (median)} & \colhead{$f_\nu$ (mean)} & \colhead{$f_\nu$ (geo.)} & {$N$}\\
    \colhead{(1)} & \colhead{(2)} & \colhead{(3)} & \colhead{(4)} & \colhead{(5)} 
}
\startdata
0.1408 & $0.1850^{+0.1638}_{-0.1019}$ & 0.1953 & 0.1621 & 108 \\
0.1424 & $0.1798^{+0.1366}_{-0.1401}$ & 0.1880 & 0.1791 & 118 \\
0.1441 & $0.1689^{+0.1320}_{-0.1058}$ & 0.1752 & 0.1618 & 117 \\
0.1457 & $0.1824^{+0.1087}_{-0.1248}$ & 0.1730 & 0.1731 & 159 \\
\enddata
\tablecomments{
Col. (1): Rest-frame wavelength in units of \micron.
Col. (2): Median composite flux density ($f_{\lambda}$) in arbitrary units. Superscripts and subscripts denote the 84th and 16th percentiles, respectively.
Col. (3): Arithmetic mean composite flux density in arbitrary units.
Col. (4): Geometric mean composite flux density in arbitrary units.
Col. (5): Number of contributing objects per spectral bin. \\
Only a portion of the spectra is displayed here to illustrate the table structure. The comprehensive dataset is provided as supplementary electronic material.
}
\end{deluxetable}

\section{Methodology}
\label{sec:method}
\subsection{Normalization}
\label{subsec:norm}
Individual spectra are normalized using a standard iterative procedure prior to constructing the final composite. We first sort the targets by redshift. For the initial pair, the weighted mean with inverse flux variance is calculated within their spectral overlap, and both spectra are normalized accordingly to create a baseline composite. We then iteratively normalize the spectrum of each subsequent target based on its overlap with the existing composite, recalculating the weighted mean at each step. This procedure is repeated for the entire sample to produce the final composite spectrum. To minimize potential systematic biases introduced during the normalization process, we independently normalize subsamples grouped by specific AGN physical properties (e.g., luminosity, black hole mass, and Eddington ratio). 

\subsection{Composite Construction}
\label{subsec:stacking}
To construct a final composite spectrum, we use a wavelength bin size that matches $R\sim 85$, approximately twice the native spectral resolution ($R\sim 35-41$) of Bands 1--4 of SPHEREx, to ensure Nyquist sampling. We note that varying the bin size does not affect the overall shape of the composite spectrum. The full spectral range approximately covers 0.14 to 4.5 $\mu$m, in which the number of contributing spectra per bin is larger than 100. Throughout the paper, we use the composite spectrum of subsamples only in spectral bins with more than 50 spectra to ensure statistical reliability.     

Stacking of individual object spectra is performed in three ways: median, mean, and geometric mean. For the geometric mean, the representative value in each spectral bin was defined as $G = \left( \prod_{i=1}^{n} x_i \right)^{\frac{1}{n}}$, where $x_i$ is the flux density within the spectral bin and $n$ is the number of associated spectral elements. We adopt the geometric mean because it preserves the shape of the continuum when it is well described by a power law, which is the case for ordinary AGNs. Prior to stacking, we apply a $3\sigma$ iterative clipping procedure within each wavelength bin to reject outliers caused by unmasked cosmic rays, instrumental artifacts, or contamination from nearby sources. This clipping procedure is also applied to median and mean composites. Throughout this process, negative flux densities are excluded so that the geometric mean is computed strictly from positive values. The scatter of the composite spectra is estimated from the 16th-84th percentile at a given spectral bin.

\section{The SPHEREx QSO Composite Spectrum}
\label{sec:results}

\subsection{Overall Spectral Shape}
\label{subsec:shape}
The SPHEREx QSO composite for the entire sample is shown in Figure~\ref{fig:fig2} and listed in Table~\ref{tab:composite}. It spans rest-frame wavelengths $0.14$--$4.5~ \mu$m, while contributing spectra are most numerous ($N>30,000$) in the range $0.4$--$2.5~\mu$m. The three composites (median, mean, and geometric mean) are consistent with each other over the full wavelength range, except at wavelengths around $\sim3~\mu$m, where the S/N is relatively low due to the higher spectral resolution. As expected, the composite spectrum clearly shows the power-law UV/optical continuum (i.e., the `Big Blue Bump'\footnote{This feature is more clearly seen with SEDs of $f_\lambda$.}) originating from the accretion disk, alongside the infrared upturn above $\sim 1~\mu$m arising from hot dust emission in the torus \cite[e.g.,][]{glikman_2006}. In addition, strong emission lines from excited hydrogen are prominently detected, while $3.3~\mu$m polycyclic aromatic hydrocarbon (PAH) emission is either absent or only weakly present.

As a comparison, we overplot the UV/optical composite spectrum of the Sloan Digital Sky Survey (SDSS) from \citet{vandenberk_2001}. To allow for a direct comparison, we rebin the SDSS composite using the spectral binning of the SPHEREx composite. Interestingly, at wavelengths below $\approx 0.4~{\mu}$m, the two composites show good agreement. This is possibly due to the similar redshift and brightness ranges of the samples used to construct the SDSS and SPHEREx composites (e.g., the median redshift of the QSO sample for the SDSS composite is 1.253, and the brightness range is $17.5 < r < 20.5$; Figure~\ref{fig:fig1} and Table~\ref{tab:sample}). However, above that wavelength, a significant discrepancy is evident. While the underlying reason for this is discussed further in Section \ref{subsec:dependence}, it may be attributed to the combination of the host galaxy contribution and its dependence on the redshift.      

To further quantify the continuum shape, we fit a power law ($f_\nu \propto \nu^{\alpha_\nu}$) to the optical and IR regions separately. For the optical band, we compute the power-law index using continuum windows at $0.15~\mu$m with a width of $0.01~\mu$m and $0.55~\mu$m with a width of $0.005~\mu$m, which are relatively free of emission lines. This yields $\alpha_{\nu, \rm opt} \approx -0.10\pm0.01$.\footnote{Throughout this study, the spectral index is measured only if there are sufficient samples ($N > 50$) per
spectral channel in the continuum windows.} The uncertainty in the slope is estimated from the series of composites generated by bootstrapping the sample for each spectral channel. We take the standard deviation of these repeated measurements as the uncertainty. We note that this value is marginally larger (i.e., harder) than the slope found for the SDSS QSO composite ($\alpha_{\nu, \rm opt} = -0.44$). This discrepancy arises partly because the SDSS composite slope was determined over a shorter wavelength range than our estimate. We adopt the longer wavelength baseline to better describe the UV/optical continuum; a power-law model anchored only at shorter wavelengths would exceed the observed flux at $\sim 0.55~\mu$m. Additionally, the slope appears to correlate with AGN properties (see Section \ref{subsec:dependence}), which may further contribute to the difference with the SDSS composite.

In the IR region, a single power law is insufficient to describe the continuum due to a ``knee'' feature around $2$--$3~\mu$m, above which a flatter power law is required to model the continuum. This feature has not been reported in previous studies and is likely a systematic effect arising from host-galaxy contamination and the luminosity dependence of the NIR SED. Specifically, the composite at longer rest-frame wavelengths relies on low-$z$ objects; these sources are less luminous and more prone to host galaxy contamination than the high-$z$ objects that dominate the shorter wavelengths (see Table~\ref{tab:sample}). Given that the NIR SED is typically flatter for low-luminosity AGNs, the broken-power-law shape appears to be a systematic artifact of these combined effects. To account for this, we estimate the power-law index using continuum windows centered at $1.2~\mu$m and $2.4~\mu$m with a width of $0.015~\mu$m. These spectral regions are selected because they are relatively free of strong emission features and minimize the bias from the systematics discussed above. This yields $\alpha_{\nu, \rm IR} = -1.46\pm0.01$. Its uncertainty is derived following the same bootstrap procedure used for $\alpha_{\nu, \rm opt}$.

\begin{figure}[t!]
\centering
\includegraphics[width=0.45\textwidth]{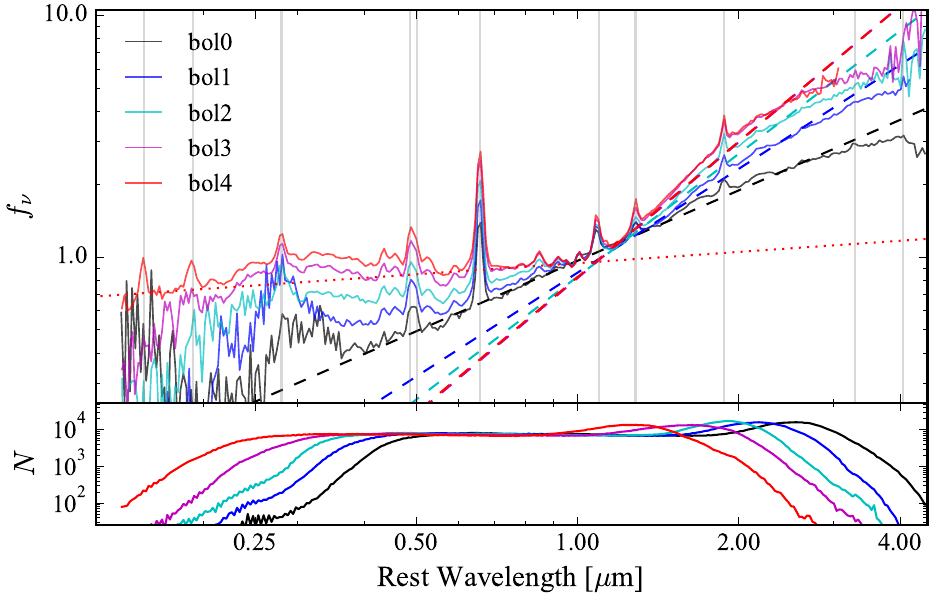}
\caption{
Same as Figure 2, but for SPHEREx composite QSO spectra grouped by bolometric luminosity ($L_{\rm bol}$). Bolometric luminosity increases from bin `bol0' through `bol4'. The composite spectrum is constructed by taking the median flux at each wavelength bin. The dotted and dashed lines denote the power-law fits to the continuum in the optical and NIR regions, respectively. The dotted line is only shown when there are sufficient samples ($N > 50$) per spectral channel in the optical bands to calculate the spectral index.
\label{fig:fig3}}
\end{figure}

\begin{figure}[ht!]
\centering
\includegraphics[width=0.45\textwidth]{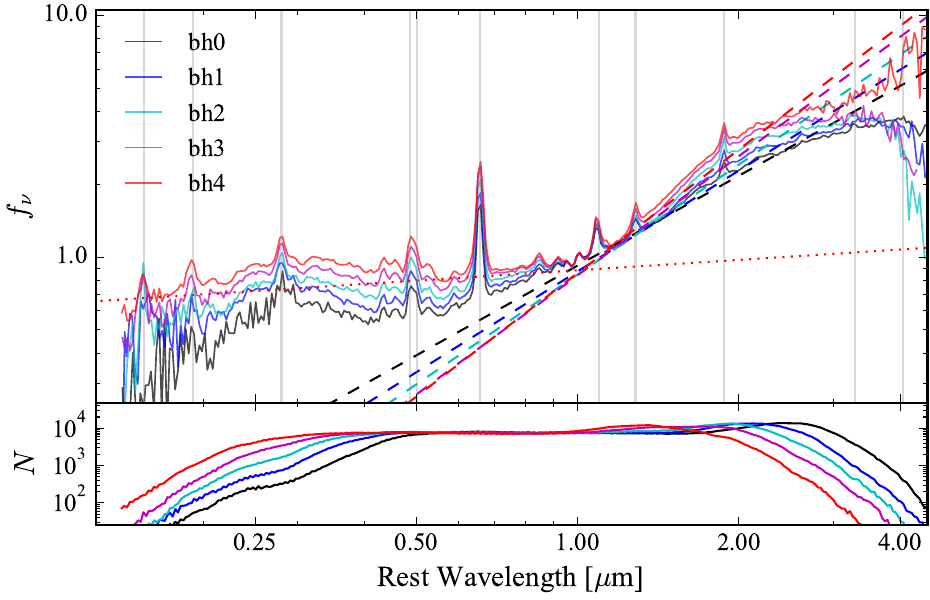}
\caption{
Same as Figure 3, but for SPHEREx composite QSO spectra grouped by BH mass. BH mass increases from bin `bh0' through `bh4'. The composite spectrum is constructed by taking the median flux at each wavelength bin.
\label{fig:fig4}}
\end{figure}

\begin{figure}[ht!]
\centering
\includegraphics[width=0.45\textwidth]{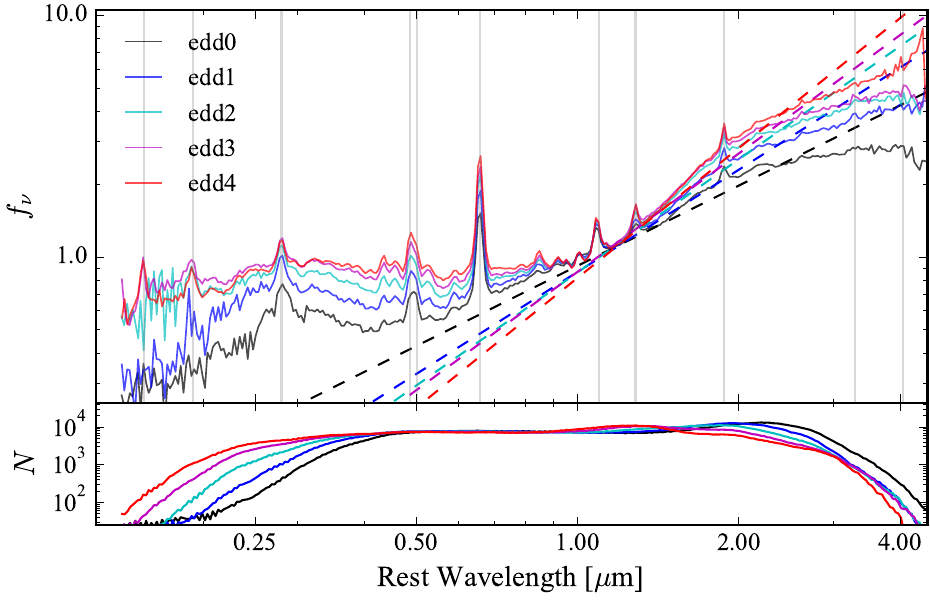}
\caption{
Same as Figure 3, but for SPHEREx composite QSO spectra grouped by Eddington ratio. Eddington ratio increases from bin `edd0' through `edd4'. The composite spectrum is constructed by taking the median flux at each wavelength bin.
\label{fig:fig5}}
\end{figure}

\subsection{Dependence on AGN properties}
\label{subsec:dependence}
\begin{figure*}[tp!]
\centering
\includegraphics[width=0.95\textwidth]{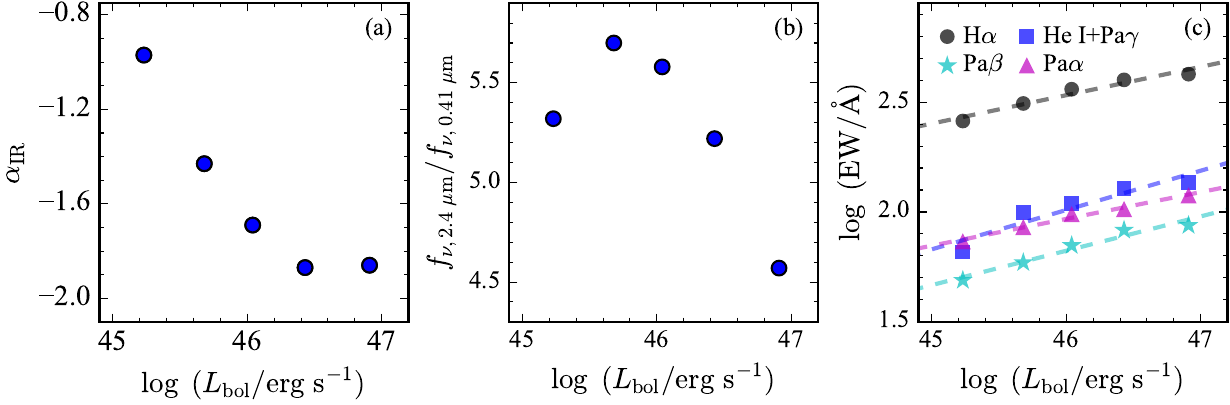}
\caption{
(a) IR spectral index, (b) flux ratio of IR to optical continuum, and (c) EWs of emission lines in the composites as a function of AGN bolometric luminosity. In the right panel, the dashed lines represent the best-fit linear relations for each emission line.
\label{fig:fig6}}
\end{figure*}

Given the sufficiently large sample size, we divide the sample into subsamples based on physical properties and construct a series of composite spectra for each group. The sub-composites binned by $L_{\rm bol}$, $M_{\rm BH}$, and $\lambda_{\rm Edd}$ are shown in Figures~\ref{fig:fig3}, \ref{fig:fig4}, and \ref{fig:fig5}, respectively. Furthermore, their composite spectra and spectral properties are summarized in the Appendix and Table~\ref{tab:spec}, respectively. Notably, across the five $L_{\rm bol}$ bins, the IR continuum slope steepens progressively from $\alpha_{\nu, {\rm IR}} = -0.96$ (bol0) to $-1.86$ (bol4). The dependence of $\alpha_{\nu, {\rm IR}}$ on the AGN luminosity is shown in Figure \ref{fig:fig6}. A similar trend is observed across BH mass and Eddington ratio bins, although the dependency is strongest for AGN luminosity and Eddington ratio. This correlation may indicate that physical properties, such as the dust covering factor, are dependent on AGN luminosity \citep[e.g.,][]{simpson_2005}.

However, it should be noted that the shape of the UV continuum also varies with AGN luminosity and the Eddington ratio. The UV continuum shape appears to be more sensitive to AGN luminosity than the NIR continuum (Figures~\ref{fig:fig3} and ~\ref{fig:fig6}). To quantify this, we compute the flux ratio of the IR to optical continuum at 2.4 and 0.41 $\mu$m, which is a primary indicator of the torus covering factor \citep[e.g.,][]{maiolino_2007}\footnote{We estimate its uncertainty in the same manner as described for $\alpha_{\nu, \rm opt}$ in Section~\ref{subsec:shape}.}. Interestingly, this ratio is inversely proportional to AGN luminosity (Figure~\ref{fig:fig6}). While the flux ratio drops for faint-end AGNs, possibly due to subtle host-galaxy contamination, these sources can be considered outliers relative to the overall trend. This finding is consistent with the receding torus model, where the covering factor decreases with increasing AGN luminosity due to dust sublimation \citep[e.g.,][]{lawrence_1991, honig_2007, son_2022, son_2023}. However, we note that some studies find no clear signature of this trend \citep[see][]{ma_2013}.

We observe a similar trend in the optical slope, though its significance cannot be definitively assessed given the current sample size. This finding is consistent with the ``bluer-when-brighter'' trend in the variability observed from ordinary AGNs, which is likely due to the positive correlation between accretion disk temperature and the accretion rate \cite[e.g.,][]{giveon_1999, wilhite_2005}. While this trend is moderately detected in $M_{\rm BH}$ and $\lambda_{\rm Edd}$ as well, these three parameters are inherently correlated. Overall, the dependency appears most prominent for $L_{\rm bol}$ and $\lambda_{\rm Edd}$, suggesting that these are the primary drivers of the observed spectral variations, consistent with theoretical predictions \cite[e.g.,][]{shakura_1973, bonning_2007}.  

\begin{figure*}[tp!]
\centering
\includegraphics[width=0.95\textwidth]{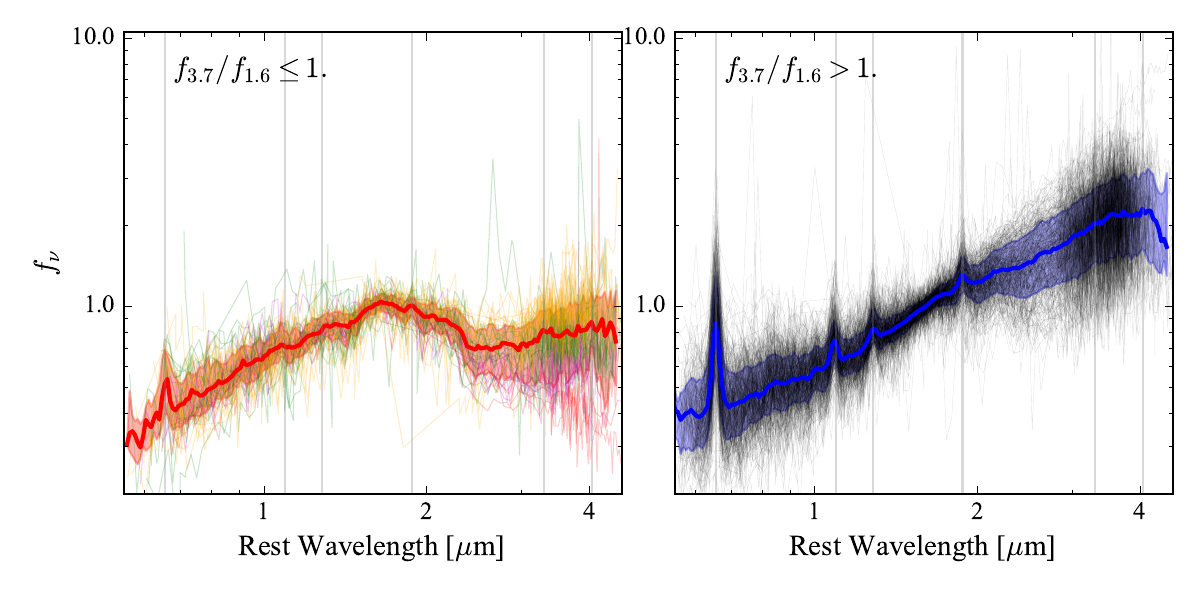}
\caption{
Examples of SPHEREx spectra for host-dominated (left) and QSO-dominated (right) objects. The classification is based on the flux ratio $f_{3.7}/f_{1.6}$, representing the relative contribution of the host galaxy versus the AGN. In the left panel, the spectra are color-coded by their $f_{3.7}/f_{1.6}$ values (orange, green, magenta, and red, in order of decreasing ratio), highlighting the transition in spectral shape as the host contribution varies. The thick red and blue lines denote the median composites for host-dominated and QSO-dominated objects, respectively. Their corresponding 16th–84th percentile ranges are represented by the shaded regions.
\label{fig:sed}}
\end{figure*}

However, we note that the dependence of the continuum slopes on AGN luminosity may be influenced by host galaxy contamination. Although we attempt to minimize host contributions through spectroscopic decomposition of the SDSS data, stellar light can remain significant in the NIR, as the stellar continuum peaks at approximately $1.6\,\mu\rm m$ in $f_\nu$. To qualitatively assess this effect, Figure~\ref{fig:sed} presents example SPHEREx spectra of low-redshift objects categorized as host-dominated or AGN-dominated based on their $1.6$-to-$3.7$ $\mu$m flux ratios estimated in the rest frame. This contribution typically steepens the optical continuum and flattens the NIR continuum, an effect particularly pronounced in low-luminosity AGNs. Consequently, we cannot rule out the possibility that the observed trends result, at least in part, from host galaxy contamination. We note that the same effect may occur in the SDSS composite from \citep{vandenberk_2001}. Because host contamination becomes more severe at longer wavelengths for a given spectral range, its impact on the SDSS composite gradually increases above 0.4 $\mu$m. Conversely, due to differences in spectral coverage, the same effect shifts to much longer wavelengths in the SPHEREx composite. This results in a significant discrepancy between the two composites from 0.4 to 0.8 $\mu$m (Figure~\ref{fig:fig2}). 

\subsection{Emission Line Properties}
\label{subsec:lines}
\begin{figure}[t!]
\centering
\includegraphics[width=0.45\textwidth]{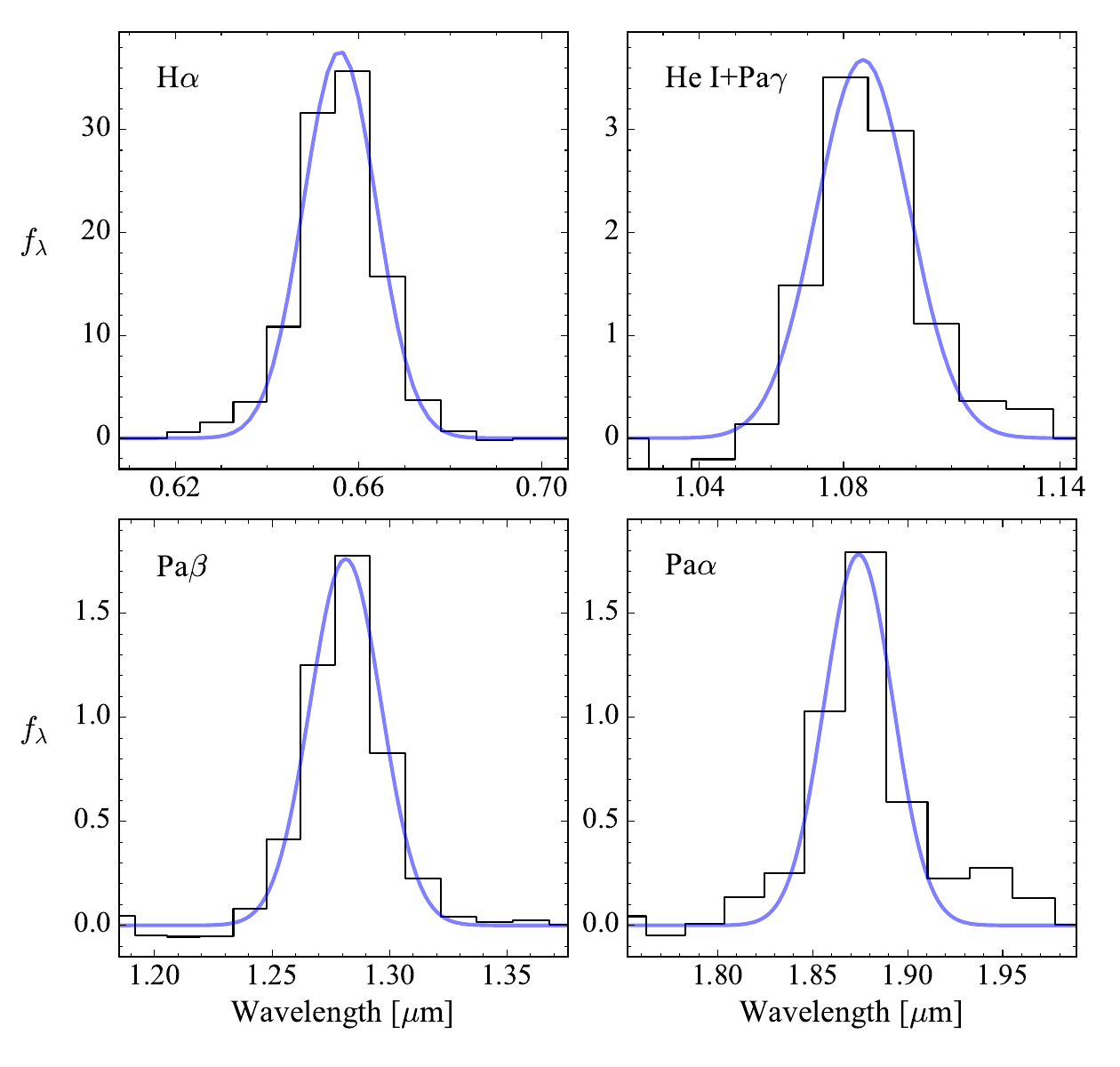}
\caption{
Examples of spectral line fits for the composite spectrum of the entire sample. The black histogram shows the observed data, while the blue line denotes the best-fit single-Gaussian model.
\label{fig:fig7}}
\end{figure}

We identify and measure the prominent emission features accessible in the SPHEREx composite: \hal, \hei+\pag\ ($1.083$~\micron), \pab\ ($1.282$~\micron), and \paa\ ($1.875$~\micron). Given the SPHEREx spectral resolution of $R \sim 35$--$130$, individual line profiles are not spectrally resolved. However, integrated line fluxes and rest-frame equivalent widths (EWs) can be measured reliably by subtracting a local continuum, modeled with a first-order polynomial, and fitting a single Gaussian to the residual profile. Example fits are shown in Figure~\ref{fig:fig7}, and the measurements for all samples are provided in Table~\ref{tab:spec}. While other Balmer lines and strong UV/optical emission lines (e.g., \ion{C}{4} and \ion{Mg}{2}) are observable, contamination from neighboring emission features and \ion{Fe}{2} multiplets makes reliable flux estimation challenging. The uncertainties in the spectral measurements are computed by varying the continuum window to fit the local baseline near the emission lines.

Additionally, the PAH 3.3 $\mu$m emission is weakly detected in the composite, indicating that its flux is intrinsically lower than that of the atomic lines. This suppression may result from the destruction of PAH molecules by high-energy photons from the accretion disk \cite[e.g.,][]{voit_1992, shim_2025}. It is in line with the fact that the PAH emission is only marginally detected in the composite of the least luminous AGNs (``bol0''; Figure ~\ref{fig:fig3}), for which the radiative destruction of PAHs is minimal. This result, however, contrasts with recent studies suggesting active star formation in QSOs (e.g., \citealp{zhuang_2020, zhuang_2021}; but see \citealp{ho_2005,kim_2006}) and the utility of PAH emission as a reliable star formation rate indicator for both star-forming galaxies and QSOs \citep[e.g.,][]{kim_2012, kim_2019, xie_2019, xie_2022}. Notably, the higher spectral resolution of Bands 5 and 6 in the SPHEREx dataset, where the 3.3 $\mu$m feature is observed for most of the sample, may degrade the S/N, contributing to the observed weakness of the emission.    

Line ratios among these emission features are instrumental in constraining the ionization mechanism and extinction within the BLR, as well as in constructing AGN templates for photometric redshift estimation and AGN identification in large-scale surveys. We therefore report line fluxes relative to the \hal\ emission. We note that the \hal\ emission is contaminated by [\ion{N}{2}] emissions. The SDSS composite spectrum yields a flux ratio of [\ion{N}{2}] emissions to the total (broad+narrow) \hal\ emission of less than $\sim1\%$, indicating that its contribution is almost negligible \citep{vandenberk_2001}. \pag\ is also heavily blended with \ion{He}{1} $\lambda 10830$, which is typically brighter than \pag. The spectral resolution of SPHEREx precludes the decomposition of these two lines; thus, our measurements represent their combined flux. For the composite spectrum for the entire sample, the Paschen-to-Balmer line ratio are $f_{\rm Pa\beta}/f_{\rm H\alpha} = 0.088$, and $f_{\rm Pa\alpha}/f_{\rm H\alpha} = 0.106$. These ratios are broadly consistent with Case B\footnote{In Case B, it is assumed that the gas is optically thick to Lyman emission due to the high gas density, consistent with the conditions in the BLR.} recombination for typical BLR conditions ($f_{\rm Pa\alpha}/f_{\rm H\alpha} = 0.11$ and  $f_{\rm Pa\beta}/f_{\rm H\alpha} = 0.10$; \citealp{kim_2010}), suggesting only weak dust reddening within the BLR on average.

Interestingly, the \paa/\hal\ flux ratio remains nearly constant across the range of sampled AGN properties. In contrast, the flux ratios of \pab\ and \ion{He}{1}+\pag\ vary significantly with both Eddington ratio and AGN luminosity. This result is consistent with recent findings suggesting that the Balmer decrement in low-$z$ AGNs is primarily governed by radiative transfer effects and excitation mechanisms rather than dust extinction \citep{son_2025}. However, this interpretation should be treated with caution, as the contribution of narrow-line emission cannot be isolated at this spectral resolution.

It is well-established that the equivalent width (EW) of the \ion{C}{4} emission line decreases with increasing AGN luminosity, a phenomenon known as the Baldwin effect \citep{baldwin_1977}. This is likely driven by a softening of the AGN SED with increasing luminosity.
Other broad emission lines, such as \ion{Mg}{2} and \hb, exhibit a similar, though less tight, trend \citep[e.g.,][]{green_2001, croom_2002}. While this effect has also been detected in narrow emission lines \citep[e.g.,][]{dietrich_2002}, results for the broad Balmer series remain controversial \citep[e.g.,][]{kovacevic_2010}. In this work, we evaluate the Baldwin effect for \hal\ and various NIR emission lines using the SPHEREx composite spectra. 

We find that the EW of H$\alpha$ is positively correlated with AGN luminosity, a trend consistently observed across all NIR transitions (Figure~\ref{fig:fig6}). More specifically, the power index $\beta$ (in the form ${\rm EW} \propto L_{\rm bol}^\beta$) derived from an ordinary least squares fit is $\sim 0.13\pm0.02, 0.16\pm0.02$, and $0.12\pm0.01$ for \hal, \pab, and \paa, respectively. Our results suggest that the standard Baldwin effect is opposite in these lines (i.e., an anti-Baldwin effect). Although our measurements integrate both narrow- and broad-line components, the narrow emission is known to exhibit a strong Baldwin effect \citep{dietrich_2002, kovacevic_2010}. Because the trend in the narrow emission is opposite to the positive correlation identified here, we conclude that the narrow-line contribution is unlikely to be the dominant driver of our observed trends. 

The anti-Baldwin effect observed in hydrogen recombination lines contrasts with that of high-ionization lines (e.g., \civ\ and \mgii). This discrepancy suggests that while the continuum shape of ionizing radiation varies in the high-energy spectral region (e.g., X-rays) in response to AGN brightness, this trend is much less pronounced for low-energy photons near the Lyman continuum, which primarily ionize hydrogen. Interestingly, the power-law index $\beta$ is similar for \hal\ and \paa, while it is marginally different for Pa$\beta$. This may indicate that the anti-Baldwin effect is driven at least partly by the underlying radiative mechanism, rather than entirely by changes in extinction or continuum shape. Nevertheless, further addressing the physical origin of the observed anti-Baldwin effect requires spectral measurements of individual objects and is beyond the scope of this study.

\begin{deluxetable*}{ccccccccccc}
\tablecaption{Spectral Properties.\label{tab:spec}}
\tablehead{
    \colhead{Sample} & \colhead{$\alpha_{\nu,{\rm opt}}$} & \colhead{$\alpha_{\nu,{\rm IR}}$} & \colhead{$f_{\rm He\ I+Pa\gamma}/f_{\rm H\alpha}$} & \colhead{$f_{\rm Pa\beta}/f_{\rm H\alpha}$} & \colhead{$f_{\rm Pa\alpha}/f_{\rm H\alpha}$} & \colhead{$\rm EW_{H\alpha}$} & \colhead{$\rm EW_{He\ I+Pa\gamma}$} & \colhead{$\rm EW_{Pa\beta}$} & \colhead{$\rm EW_{Pa\alpha}$} & \colhead{$f_{\rm IR}/f_{\rm opt}$}\\
    \colhead{(1)} & \colhead{(2)} & \colhead{(3)} & \colhead{(4)} & \colhead{(5)} & \colhead{(6)} & \colhead{(7)} & \colhead{(8)} & \colhead{(9)} & \colhead{(10)} & \colhead{(11)}  
}
\startdata
All & $-0.10 \pm 0.01$ & $-1.46 \pm 0.01$ & $0.1576\pm0.0017$ & $0.0885\pm0.0002$ & $0.1064\pm0.0009$ & $347.61\pm1.37$ & $112.16\pm1.64$ & $72.24\pm0.12$ & $95.49\pm1.15$ & $4.32 \pm 0.02$ \\
bol0 & \nd & $-0.97 \pm 0.01$ & $0.1611\pm0.0004$ & $0.0999\pm0.0004$ & $0.1049\pm0.0011$ & $259.33\pm1.53$ & $65.85\pm0.13$ & $48.92\pm0.29$ & $73.56\pm1.33$ & $5.32 \pm 0.10$ \\
bol1 & \nd & $-1.43 \pm 0.01$ & $0.1775\pm0.0028$ & $0.0884\pm0.0003$ & $0.1092\pm0.0016$ & $311.63\pm0.63$ & $99.82\pm1.40$ & $58.90\pm0.19$ & $85.42\pm1.09$ & $5.70 \pm 0.05$ \\
bol2 & \nd & $-1.69 \pm 0.01$ & $0.1523\pm0.0014$ & $0.0846\pm0.0003$ & $0.1137\pm0.0011$ & $361.78\pm0.56$ & $109.68\pm1.44$ & $70.43\pm0.29$ & $97.85\pm1.37$ & $5.58 \pm 0.04$ \\
bol3 & \nd & $-1.87 \pm 0.01$ & $0.1429\pm0.0019$ & $0.0826\pm0.0002$ & $0.1068\pm0.0012$ & $399.01\pm0.63$ & $127.60\pm2.05$ & $82.68\pm0.24$ & $103.17\pm1.43$ & $5.22 \pm 0.04$ \\
bol4 & $-0.15 \pm 0.01$ & $-1.86 \pm 0.02$ & $0.1357\pm0.0032$ & $0.0787\pm0.0003$ & $0.1109\pm0.0027$ & $423.56\pm1.90$ & $135.84\pm2.12$ & $86.96\pm0.14$ & $119.09\pm1.84$ & $4.57 \pm 0.05$ \\
bh0 & \nd & $-1.23 \pm 0.01$ & $0.1641\pm0.0007$ & $0.0929\pm0.0002$ & $0.1125\pm0.0009$ & $283.56\pm0.95$ & $81.00\pm0.58$ & $53.42\pm0.13$ & $82.00\pm1.11$ & $5.03 \pm 0.06$ \\
bh1 & \nd & $-1.39 \pm 0.01$ & $0.1659\pm0.0011$ & $0.0952\pm0.0002$ & $0.1069\pm0.0009$ & $318.68\pm0.98$ & $99.42\pm1.08$ & $67.54\pm0.20$ & $86.89\pm1.19$ & $4.83 \pm 0.04$ \\
bh2 & \nd & $-1.51 \pm 0.01$ & $0.1516\pm0.0015$ & $0.0836\pm0.0003$ & $0.1064\pm0.0010$ & $356.48\pm0.63$ & $109.04\pm1.58$ & $70.32\pm0.39$ & $94.83\pm1.35$ & $4.61 \pm 0.03$ \\
bh3 & \nd & $-1.64 \pm 0.01$ & $0.1528\pm0.0010$ & $0.0820\pm0.0001$ & $0.1072\pm0.0013$ & $388.55\pm0.44$ & $129.06\pm1.15$ & $78.21\pm0.16$ & $103.74\pm1.54$ & $4.58 \pm 0.04$ \\
bh4 & $-0.14 \pm 0.01$ & $-1.70 \pm 0.02$ & $0.1461\pm0.0025$ & $0.0823\pm0.0001$ & $0.1163\pm0.0015$ & $405.82\pm0.89$ & $133.53\pm2.58$ & $82.89\pm0.14$ & $117.76\pm1.67$ & $4.38 \pm 0.05$ \\
edd0 & \nd & $-1.10 \pm 0.01$ & $0.1722\pm0.0013$ & $0.0940\pm0.0004$ & $0.1191\pm0.0015$ & $298.36\pm1.33$ & $87.64\pm1.04$ & $57.70\pm0.33$ & $94.30\pm1.82$ & $4.89 \pm 0.05$ \\
edd1 & \nd & $-1.40 \pm 0.01$ & $0.1648\pm0.0014$ & $0.0924\pm0.0003$ & $0.1025\pm0.0009$ & $342.93\pm0.67$ & $105.99\pm1.33$ & $70.02\pm0.32$ & $87.74\pm1.14$ & $4.83 \pm 0.04$ \\
edd2 & \nd & $-1.55 \pm 0.01$ & $0.1543\pm0.0013$ & $0.0833\pm0.0002$ & $0.1060\pm0.0016$ & $355.99\pm0.80$ & $112.60\pm0.91$ & $68.85\pm0.16$ & $93.09\pm1.29$ & $4.64 \pm 0.03$ \\
edd3 & \nd & $-1.63 \pm 0.01$ & $0.1465\pm0.0020$ & $0.0857\pm0.0001$ & $0.1041\pm0.0011$ & $381.22\pm1.04$ & $119.70\pm1.99$ & $77.72\pm0.11$ & $97.72\pm1.34$ & $4.42 \pm 0.03$ \\
edd4 & \nd & $-1.78 \pm 0.01$ & $0.1398\pm0.0016$ & $0.0813\pm0.0003$ & $0.0964\pm0.0017$ & $396.91\pm1.06$ & $129.35\pm1.58$ & $84.76\pm0.31$ & $98.13\pm1.57$ & $4.51 \pm 0.02$ \\
\enddata
\tablecomments{
Col. (1): Subsample.
Col. (2): Optical power-law index ($\alpha_{\nu, \text{opt}}$) where $f_\nu \propto \nu^{\alpha}$, calculated between $0.15$ and $0.55$ \micron. This value is only listed if the number of samples in two spectral channels is greater than 50.
Col. (3): NIR power-law index ($\alpha_{\nu, {\rm IR}}$) calculated between $1.2$ and $2.4$ \micron.
Col. (4): Flux ratio of \hei\ and \pag\ to \hal.  
Col. (5): Flux ratio of \pab\ to \hal.
Col. (6): Flux ratio of \paa\ to \hal.
Col. (7): Equivalent width of \hal\ emission in units of \AA.
Col. (8): Equivalent width of \hei\ and \pag\ emissions in units of \AA.
Col. (9): Equivalent width of \pab\ emission in units of \AA.
Col. (10): Equivalent width of \paa\ emission in units of \AA.
Col. (11): Flux ratio [${f_{\nu}(2.4\,{\rm \mu m})}/{f_{\nu}(0.41\,{\rm \mu m})}$] of IR to optical continuum, measured at $2.4$ \micron, and $0.41$ \micron\, respectively. 
}
\end{deluxetable*}

\section{Discussion}
\label{sec:discussion}

\subsection{Comparison with Previous NIR Composite Spectra}
\label{subsec:comparison}
\begin{figure}[t!]
\centering
\includegraphics[width=0.45\textwidth]{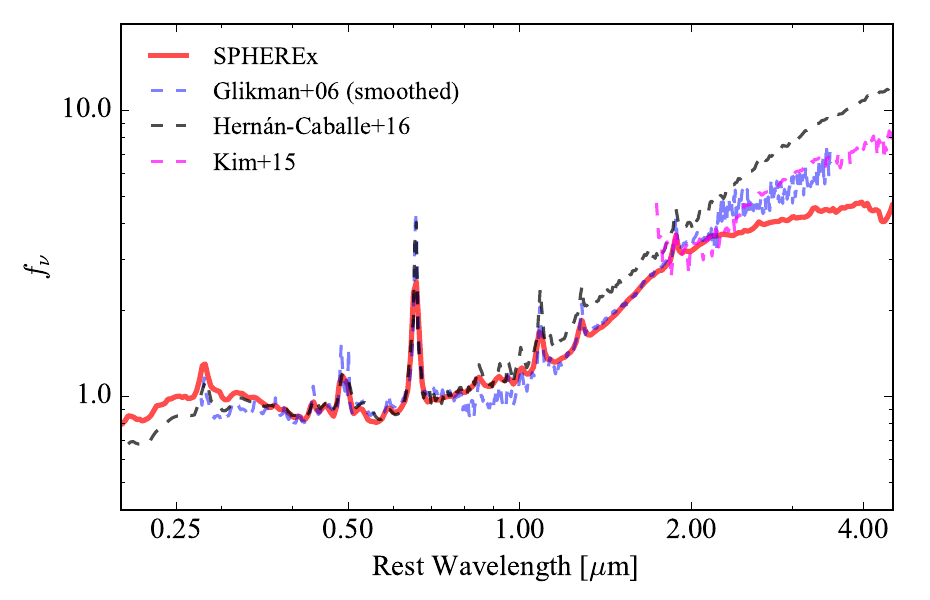}
\caption{
Comparison of NIR QSO composite spectra with previous studies. While the red line denotes the composite spectrum from this study, blue, black, and magenta dashed lines denote the composites from \citet{glikman_2006}, \citet{hernan-Caballe_2016}, and \citet{kim_2015}, respectively.
\label{fig:fig8}}
\end{figure}

To assess the quality and reliability of the SPHEREx composite constructed in this study, we compare it with previously published NIR QSO composites \citep{glikman_2006, kim_2015, hernan-Caballe_2016, kimd_2026}, as shown in Figure~\ref{fig:fig8}. In the overlapping wavelength range ($\sim 0.9$--$2.5$~\micron), the four composites agree qualitatively in the overall shape of the continuum. However, quantitative differences among the composites are pronounced in the NIR continuum slope.

This variation is somewhat anticipated given the observed dependence of the continuum on AGN luminosity. The composites from \citet{glikman_2006} and \citet{kim_2015} agree well with each other in the NIR because both are based on relatively low-redshift  AGNs with moderate luminosities. In contrast, \citet{hernan-Caballe_2016} constructed composites including luminous, high-$z$ AGNs, resulting in a steeper NIR slope compared to the other works (see also \citealt{selsing_2016}).

Furthermore, previous studies generally excluded low-luminosity AGNs from their analyses. This suggests that their host galaxy contribution is minimal, leading to a lack of a ``knee'' feature in the NIR continuum. This finding is consistent with our observation that our composite for high-luminosity objects is in good agreement with that of \citet{hernan-Caballe_2016}. This assessment reveals that constructing a composite spectrum requires careful treatment to control for AGN properties and redshift distributions to match the intended use.

\subsection{Applications}
\label{subsec:applications}

One of the primary motivations for constructing this composite is its application as a spectral template for the photometric redshift estimation of QSOs, particularly for the SPHEREx all-sky survey and forthcoming datasets. The low-resolution spectral information from SPHEREx channels encodes redshift-dependent color patterns from broad emission lines that can be exploited during photo-$z$ fitting. Incorporating an empirically derived composite, rather than a theoretical or sparsely sampled prior template, is expected to enhance the reliability of these estimations. Furthermore, the availability of sub-composites binned by physical parameters may improve photo-$z$ performance, since emission lines and continuum shapes change significantly with AGN parameters, as shown in this study and previous works \cite[e.g.,][]{boroson_1992, shen_2014}. Quantitative assessments of photo-$z$ recovery using this composite will be presented in a future study.

Additionally, this dataset enables robust QSO identification when combined with templates of inactive galaxies. Historically, QSO identification has been conducted using color information in the optical and MIR regimes. However, such selection is often performed using color–color diagrams without detailed SED fitting, which can introduce systematic biases in AGN selection \cite[e.g.,][]{kim_2026}. With our dataset, a more sophisticated QSO selection, incorporating counterpart SEDs from inactive galaxies and accounting for the variance in AGN SEDs across a broad range of properties, is now possible.

\subsection{Caveats and Limitations}
\label{subsec:caveats}
Because of the unique characteristics of the SPHEREx dataset, several caveats apply to the use of these composites:

\begin{itemize}
    \item \textbf{Spectral resolution.} Because the SPHEREx LVF channels ($R \sim 35$--130) cannot resolve individual line profiles, emission lines are largely unresolved and often blended with other features, such as Fe multiplets. In addition, as described in Section~\ref{subsec:lines}, we cannot reliably separate the emission lines into broad and narrow components. Consequently, spectral measurements of individual emission lines should be interpreted with caution. 

    \item \textbf{Variability.} Due to the characteristics of the LVF filters and the SPHEREx survey strategy, the spectrum of an individual target is typically obtained over a month on average. Because AGNs vary intrinsically on timescales of days to years, this asynchronous sampling could introduce systematic biases in the measured spectral shape. Accounting for cosmic time dilation, the corresponding rest-frame observing timescale is $\sim 15$ days. Since the variability amplitude of QSOs typically peaks on year-long timescales, the expected variability should be $< 0.1$~mag and becomes significantly smaller in the NIR bands \citep[e.g.,][]{kim_2024}. Therefore, the overall effect of variability is likely negligible, less than by $\sim 10$\%.

    \item \textbf{Host-galaxy contamination.} Host-galaxy contamination significantly alters the SED of the target AGN across a wide spectral range. This effect is more prominent in low-luminosity and low-redshift AGNs. When constructing composite spectra over broad wavelength ranges, this contamination introduces nonlinear effects along the wavelength axis and systematic biases in the spectral shape. Users should therefore exercise extreme care when utilizing these composites in regimes where host contribution is non-negligible. One possible way to mitigate this effect is to use the composites from the high-luminosity AGN, where the host contribution is minimal. 
    
\end{itemize}

\section{Summary and Conclusions}
\label{sec:summary}

We construct a composite QSO spectrum from SPHEREx all-sky survey spectrophotometry of $\sim61,000$ objects. Our main findings are:

\begin{enumerate}
    \item The SPHEREx QSO composite spans rest-frame wavelengths $0.14$--$4.5~\mu\text{m}$, providing the first unified UV-to-NIR QSO SED template from the all-sky survey. The median, mean, and geometric-mean composites are mutually consistent across the full wavelength range, except at $> 3~\mu{\rm m}$.

    \item The UV/optical continuum is well described by a power law with $\alpha_{\nu, \text{opt}} = -0.10$, which is marginally harder than the SDSS composite value of $-0.44$, partly due to the longer wavelength baseline adopted in this work. In the NIR, the continuum steepens to $\alpha_{\nu, \text{IR}} = -1.46$, reflecting a dominant hot-dust contribution from the AGN torus at $\lambda \gtrsim 1~\mu\text{m}$. Both slopes depend on AGN properties, with more luminous AGNs exhibiting flatter UV/optical and steeper NIR continua.

    \item The ratio of IR to optical flux appears to decrease with increasing luminosity. This trend suggests that the dust-covering factor may be inversely proportional to the AGN luminosity, a result consistent with the ``receding torus'' model \citep[e.g.,][]{lawrence_1991}. However, this finding remains inconclusive due to the limited sample size of the current dataset.   

    \item Broad emission lines, including H$\alpha$, \ion{He}{1}+\pag, \pab, and \paa, are clearly detected. Their Paschen-to-Balmer line ratios ($f_{\rm Pa\beta}/f_{\rm H\alpha} = 0.088$ and $f_{\rm Pa\alpha}/f_{\rm H\alpha} = 0.106$) are in good agreement with Case B recombination predictions, suggesting that internal extinction within the BLR is almost negligible on average.

    \item The equivalent widths of H$\alpha$ and all detected NIR emission lines increase with AGN luminosity, contrary to the trend expected from the classical Baldwin effect. This positive correlation is consistently observed across all NIR transitions and is unlikely to be driven by the narrow-line component, which is known to exhibit the opposite trend.

    \item The shape of the composite is sensitive to host-galaxy contamination, particularly at NIR wavelengths. This contamination is most pronounced in low-luminosity and low-redshift objects, manifesting as a "knee" feature in the NIR continuum at approximately $2-3$ \micron. We note that this composite should be used with caution.
\end{enumerate}

\begin{acknowledgments}
We are grateful to an anonymous referee for insightful comments that greatly improved the manuscript. We thank Luis Ho for the insightful discussion. This work was supported by the National Research Foundation of Korea (NRF) grant funded by the Korean government (MSIT) (Nos. RS-2024-00347548 and RS-2025-16066624) and the Yonsei University Research Fund of 2025 (2025-22-0402). Y. K. was supported by the National Research Foundation of Korea (NRF) grant funded by the Korean government (MSIT) (No. RS-2026-25476464). Part of the research described in this paper was carried out at the Jet Propulsion Laboratory, California Institute of Technology, under a contract with the National Aeronautics and Space Administration (80NM0018D0004). K.O. acknowledges support from the National Research Foundation of Korea (NRF) grant funded by the Korea government (MSIT) (RS-2025-00553982). This work utilizes the high-speed network connections provided by the Korea Research Environment Open NETwork (KREONET), which is managed and operated by the Korea Institute of Science and Technology Information (KISTI), and the computing resources of the Institute for Basic Science (IBS) Research Solution Center (RSC).

This publication makes use of data products from the Spectro-Photometer for the History of the Universe, Epoch of Reionization and Ices Explorer (SPHEREx), which is a joint project of the Jet Propulsion Laboratory and the California Institute of Technology, and is funded by the National Aeronautics and Space Administration.

Funding for the Sloan Digital Sky Survey V has been provided by the Alfred P. Sloan Foundation, the Heising-Simons Foundation, the National Science Foundation, and the Participating Institutions. SDSS acknowledges support and resources from the Center for High-Performance Computing at the University of Utah. SDSS telescopes are located at Apache Point Observatory, funded by the Astrophysical Research Consortium and operated by New Mexico State University, and at Las Campanas Observatory, operated by the Carnegie Institution for Science. The SDSS website is \url{www.sdss.org}.

SDSS is managed by the Astrophysical Research Consortium for the Participating Institutions of the SDSS Collaboration, including the Carnegie Institution for Science, Chilean National Time Allocation Committee (CNTAC) ratified researchers, Caltech, the Gotham Participation Group, Harvard University, Heidelberg University, The Flatiron Institute, The Johns Hopkins University, L'Ecole polytechnique f\'{e}d\'{e}rale de Lausanne (EPFL), Leibniz-Institut f\"{u}r Astrophysik Potsdam (AIP), Max-Planck-Institut f\"{u}r Astronomie (MPIA Heidelberg), Max-Planck-Institut f\"{u}r Extraterrestrische Physik (MPE), Nanjing University, National Astronomical Observatories of China (NAOC), New Mexico State University, The Ohio State University, Pennsylvania State University, Smithsonian Astrophysical Observatory, Space Telescope Science Institute (STScI), the Stellar Astrophysics Participation Group, Universidad Nacional Aut\'{o}noma de M\'{e}xico, University of Arizona, University of Colorado Boulder, University of Illinois at Urbana-Champaign, University of Toronto, University of Utah, University of Virginia, Yale University, and Yunnan University.

\end{acknowledgments}

\facilities{SPHEREx, Sloan}

\software{
    Astropy \citep{astropy_2013, astropy_2018, astropy_2022},
    SciPy \citep{scipy_2020}
}

\bibliography{ms}{}
\bibliographystyle{aasjournalv7}

\appendix

\section{Median Composites for Subsamples}
Here, we present the median SPHEREx composites for subsamples divided by AGN bolometric luminosity (Table~\ref{tab:composite_sub1}), BH mass (Table~\ref{tab:composite_sub2}), and Eddington ratio (Table~\ref{tab:composite_sub3}). 

\restartappendixnumbering

\begin{deluxetable*}{cccccccccc}
\tablecaption{Median SPHEREx QSO Composite Spectra for Subsamples Divided by $L_{\rm bol}$.\label{tab:composite_sub1}}
\tablehead{
    \multicolumn2c{bol0} & \multicolumn2c{bol1} & \multicolumn2c{bol2} & \multicolumn2c{bol3} & \multicolumn2c{bol4} \\
    \cmidrule(lr){1-2}  \cmidrule(lr){3-4} 
    \cmidrule(lr){5-6}  \cmidrule(lr){7-8}
    \cmidrule(lr){9-10} 
    \colhead{$\lambda_{\rm rest}$} & \colhead{$f_\nu$} &
    \colhead{$\lambda_{\rm rest}$} & \colhead{$f_\nu$} &
    \colhead{$\lambda_{\rm rest}$} & \colhead{$f_\nu$} &
    \colhead{$\lambda_{\rm rest}$} & \colhead{$f_\nu$} &
    \colhead{$\lambda_{\rm rest}$} & \colhead{$f_\nu$} 
}
\startdata
0.2420 & $0.1056^{+0.2662}_{-0.0955}$ & 0.2232 & $0.3053^{+0.2829}_{-0.2491}$ & 0.1967 & $0.1793^{+0.4080}_{-0.1274}$ & 0.1751 & $0.1973^{+0.2741}_{-0.1162}$ & 0.1408 & $0.3851^{+0.2962}_{-0.1977}$ \\
0.2842 & $0.2752^{+0.3686}_{-0.2153}$ & 0.2258 & $0.3445^{+0.7169}_{-0.2825}$ & 0.2014 & $0.2027^{+0.2779}_{-0.1463}$ & 0.1771 & $0.1531^{+0.3335}_{-0.1057}$ & 0.1424 & $0.4377^{+0.1874}_{-0.2726}$ \\
0.2910 & $0.2094^{+0.3302}_{-0.1804}$ & 0.2311 & $0.4595^{+0.4730}_{-0.4114}$ & 0.2035 & $0.2406^{+0.3198}_{-0.1488}$ & 0.1793 & $0.2004^{+0.3070}_{-0.1342}$ & 0.1441 & $0.3783^{+0.1980}_{-0.1938}$ \\
0.2941 & $0.2647^{+0.3630}_{-0.2071}$ & 0.2337 & $0.2672^{+0.4791}_{-0.2454}$ & 0.2060 & $0.1979^{+0.3452}_{-0.1458}$ & 0.1813 & $0.2533^{+0.2687}_{-0.1624}$ & 0.1457 & $0.3818^{+0.2152}_{-0.2195}$ \\
\enddata
\tablecomments{
Only a portion of the spectra is displayed here to illustrate the table structure. The comprehensive dataset is provided as supplementary electronic material.
}
\end{deluxetable*}

\begin{deluxetable*}{cccccccccc}
\tablecaption{Median SPHEREx QSO Composite Spectra for Subsamples Divided by $M_{\rm BH}$.\label{tab:composite_sub2}}
\tablehead{
    \multicolumn2c{bh0} & \multicolumn2c{bh1} & \multicolumn2c{bh2} & \multicolumn2c{bh3} & \multicolumn2c{bh4} \\
    \cmidrule(lr){1-2}  \cmidrule(lr){3-4} 
    \cmidrule(lr){5-6}  \cmidrule(lr){7-8}
    \cmidrule(lr){9-10} 
    \colhead{$\lambda_{\rm rest}$} & \colhead{$f_\nu$} &
    \colhead{$\lambda_{\rm rest}$} & \colhead{$f_\nu$} &
    \colhead{$\lambda_{\rm rest}$} & \colhead{$f_\nu$} &
    \colhead{$\lambda_{\rm rest}$} & \colhead{$f_\nu$} &
    \colhead{$\lambda_{\rm rest}$} & \colhead{$f_\nu$} 
}
\startdata
0.1898 & $0.2204^{+0.3529}_{-0.1174}$ & 0.1732 & $0.1719^{+0.1782}_{-0.1092}$ & 0.1693 & $0.2118^{+0.1080}_{-0.1443}$ & 0.1545 & $0.3195^{+0.1913}_{-0.1986}$ & 0.1408 & $0.2353^{+0.1763}_{-0.1230}$ \\
0.1921 & $0.2063^{+0.2539}_{-0.1585}$ & 0.1773 & $0.2559^{+0.1399}_{-0.1807}$ & 0.1713 & $0.1751^{+0.2230}_{-0.0995}$ & 0.1581 & $0.2029^{+0.1429}_{-0.0987}$ & 0.1424 & $0.2532^{+0.1024}_{-0.1700}$ \\
0.1941 & $0.1568^{+0.2699}_{-0.1713}$ & 0.1792 & $0.1351^{+0.1637}_{-0.0883}$ & 0.1733 & $0.1987^{+0.1272}_{-0.1226}$ & 0.1599 & $0.2451^{+0.1502}_{-0.0981}$ & 0.1440 & $0.2173^{+0.1264}_{-0.1292}$ \\
0.1965 & $0.1847^{+0.1962}_{-0.1643}$ & 0.1813 & $0.2012^{+0.2660}_{-0.1343}$ & 0.1753 & $0.1956^{+0.1468}_{-0.1187}$ & 0.1618 & $0.2147^{+0.1955}_{-0.0985}$ & 0.1457 & $0.2191^{+0.1292}_{-0.1277}$ \\
\enddata
\tablecomments{
Only a portion of the spectra is displayed here to illustrate the table structure. The comprehensive dataset is provided as supplementary electronic material.
}
\end{deluxetable*}

\begin{deluxetable*}{cccccccccc}
\tablecaption{Median SPHEREx QSO Composite Spectra for Subsamples Divided by $\lambda_{\rm Edd}$.\label{tab:composite_sub3}}
\tablehead{
    \multicolumn2c{edd0} & \multicolumn2c{edd1} & \multicolumn2c{edd2} & \multicolumn2c{edd3} & \multicolumn2c{edd4} \\
    \cmidrule(lr){1-2}  \cmidrule(lr){3-4} 
    \cmidrule(lr){5-6}  \cmidrule(lr){7-8}
    \cmidrule(lr){9-10} 
    \colhead{$\lambda_{\rm rest}$} & \colhead{$f_\nu$} &
    \colhead{$\lambda_{\rm rest}$} & \colhead{$f_\nu$} &
    \colhead{$\lambda_{\rm rest}$} & \colhead{$f_\nu$} &
    \colhead{$\lambda_{\rm rest}$} & \colhead{$f_\nu$} &
    \colhead{$\lambda_{\rm rest}$} & \colhead{$f_\nu$} 
}
\startdata
0.1989 & $0.1548^{+0.2068}_{-0.1152}$ & 0.1946 & $0.1867^{+0.1850}_{-0.1062}$ & 0.1772 & $0.3058^{+0.3245}_{-0.2012}$ & 0.1579 & $0.2634^{+0.1667}_{-0.1322}$ & 0.1410 & $0.2650^{+0.1792}_{-0.1470}$ \\
0.2034 & $0.1894^{+0.1210}_{-0.1262}$ & 0.1966 & $0.1769^{+0.2391}_{-0.1588}$ & 0.1794 & $0.4465^{+0.3011}_{-0.2761}$ & 0.1599 & $0.3019^{+0.1847}_{-0.1468}$ & 0.1442 & $0.2484^{+0.1622}_{-0.1479}$ \\
0.2082 & $0.1991^{+0.1694}_{-0.1175}$ & 0.1991 & $0.2030^{+0.2003}_{-0.1418}$ & 0.1813 & $0.4649^{+0.3112}_{-0.3000}$ & 0.1616 & $0.3148^{+0.1390}_{-0.1852}$ & 0.1458 & $0.2724^{+0.2024}_{-0.1745}$ \\
0.2131 & $0.2094^{+0.1284}_{-0.1307}$ & 0.2012 & $0.1528^{+0.2763}_{-0.0728}$ & 0.1835 & $0.3986^{+0.3116}_{-0.1564}$ & 0.1635 & $0.2752^{+0.1738}_{-0.1339}$ & 0.1475 & $0.2882^{+0.1875}_{-0.1401}$ \\
\enddata
\tablecomments{
Only a portion of the spectra is displayed here to illustrate the table structure. The comprehensive dataset is provided as supplementary electronic material.
}
\end{deluxetable*}

\end{document}